\documentclass[aps,prl,twocolumn,showpacs,superscriptaddress,groupedaddress]{revtex4}

\usepackage{amsmath,amsfonts}
\usepackage{bm}
\usepackage{graphicx}
\usepackage[pdftex,usenames,dvipsnames]{color}
\usepackage{comment}

\def\opone{\leavevmode\hbox{\small1\kern-3.8pt\normalsize1}}

\begin{document}
\title{Tm$^{3+}$:Y$_3$Ga$_5$O$_{12}$ materials for spectrally multiplexed quantum memories}
\author{C. W. Thiel}
\affiliation{Department of Physics, Montana State University, Bozeman, Montana 59717, USA}
\author{N. Sinclair}
\affiliation{Institute for Quantum Science and Technology, and Department of Physics \& Astronomy, University of Calgary, Calgary, Alberta T2N 1N4, Canada}
\author{W. Tittel}
\affiliation{Institute for Quantum Science and Technology, and Department of Physics \& Astronomy, University of Calgary, Calgary, Alberta T2N 1N4, Canada}
\author{R. L. Cone}
\affiliation{Department of Physics, Montana State University, Bozeman, Montana 59717, USA}

\begin{abstract}
We investigate the relevant spectroscopic properties of the 795 nm $^3$H$_6$$\leftrightarrow$$^3$H$_4$ transition in 1\% Tm$^{3+}$:Y$_3$Ga$_5$O$_{12}$ at temperatures as low as 1.2 K for optical quantum memories based on persistent spectral tailoring of narrow absorption features. Our measurements reveal that this transition has uniform coherence properties over a 56 GHz  bandwidth, and a simple hyperfine structure split by $\pm$44 MHz/T with lifetimes of up to hours. Furthermore, we find a $^3$F$_4$ population lifetime of 64 ms -- one of the longest lifetimes observed for an electronic level in a solid -- and an exceptionally long coherence lifetime of 490 $\mu$s -- the longest ever observed for optical transitions of Tm$^{3+}$ ions in a crystal. Our results suggest that this material allows realizing broadband quantum memories that enable spectrally multiplexed quantum repeaters.
\end{abstract}
 
\maketitle

Optical quantum memories relying on reversible mapping of quantum states between photons and matter constitute key ingredients for long-distance quantum communication via quantum repeaters \cite{sangouard2011a,lvovsky2009a}. Many approaches to quantum memory are being pursued \cite{lvovsky2009a,bussieres2013a}; here we focus on quantum state storage based on atomic frequency combs (AFCs) \cite{afzelius2009a} and spectral multiplexing \cite{sinclair2014a} as this approach offers a clear, and possibly the simplest, path towards meeting the stringent demands for quantum repeaters. For the repeater architecture proposed by Sinclair et al. \cite{sinclair2014a}, these demands are storage times measured in hundreds of microseconds, simultaneous storage of around 1000 photons, feed-forward-based retrieval of a particular photon in a predefined mode, storage efficiency and fidelity approaching unity, and a bandwidth per spectral channel on the order of 100 MHz to GHz.

Rare-earth-ion doped (RE) crystals cooled to cryogenic temperatures are promising candidates for creating spectrally multiplexed AFC quantum memories due to their favorable properties \cite{tittel2010a}. These properties include optical coherence lifetimes exceeding milliseconds and second-long spin coherence lifetimes, either of which allows achieving the necessary storage time. These systems also offer hour-long population lifetimes in auxiliary states needed to create the required AFCs through frequency-selective persistent spectral hole burning. Furthermore, inhomogeneously broadened absorption linewidths of hundreds of GHz and hyperfine splittings of up to GHz promise the generation of the appropriate number of spectrally-multiplexed memory channels. Finally, the use of a solid-state material facilitates incorporation into impedance matched cavities and allows integration with optical waveguides \cite{sinclair2010a,sabooni2013a}. Despite these promising attributes and accelerated experimental progress \cite{bussieres2013a, sinclair2014a, sabooni2013a, riedmatten2008a, afzelius2010a, hedges2010a, clausen2011a, saglamyurek2011a,jin2013a,rielander2014a,jobez2014a}, no RE material (or any other system) has yet allowed constructing optical quantum memories suitable for use in a quantum repeater.

Here we introduce a promising new RE quantum memory material Tm$^{3+}$:Y$_3$Ga$_5$O$_{12}$ (Tm:YGG) and report properties of the inhomogeneously broadened $^3$H$_6$$\leftrightarrow$$^3$H$_4$ transition at 795 nm for temperatures as low as 1.2 K. Very little has been previously reported on either the spectral hole burning or coherence properties of this material; however, promising initial measurements of $1/e$ coherence lifetimes as long as 410 $\mu$s \cite{sun2005a,thiel2011a} motived the detailed studies relevant for spectrally multiplexed AFC-based quantum memories that are presented here. Using photon echo and spectral hole burning measurements \cite{macfarlane1987a}, we find uniform coherence properties over the entire 56 GHz inhomogeneous linewidth, as well as suitable excited and hyperfine level structure and lifetimes. This suggests that, in conjunction with impedance-matched cavities \cite{afzelius2010b,moiseev2010a}, Tm:YGG is the first, or, at least, the first Tm$^{3+}$-doped, material that satisfies all properties required for the AFC-based spectrally multiplexed quantum repeater proposed in Ref. \cite{sinclair2014a}. Furthermore, we note that our results are applicable to other memory approaches based on persistent spectral tailoring \cite{tittel2010a}.

\noindent
\textbf{\textit{Experimental methods--}} Measurements are carried out on an optically transparent, 1 cm long single crystal of 1\%Tm:YGG grown by Scientific Materials Corp. (Bozeman, MT) with the laser propagating along a $<$110$>$ direction, and the applied magnetic field and linear optical polarization both set parallel to an orthogonal $<$111$>$ axis. Our sample is cooled either by immersion in superfluid liquid helium or by a constant flow of helium exchange gas within a cryostat. Magnetic fields of up to 500 G are applied using a Helmholtz coil, while a field of 0.64 T is generated by mounting the crystal between a pair of NdFeB block magnets. A continuous-wave Ti:Sapphire ring laser with a linewidth of $\sim$100 kHz and output power of typically 425 mW is used as the light source with the frequency monitored using a wavemeter. A pair of acousto-optic modulators (AOMs) are used in series to generate pulses for echo and hole burning measurements. Spectral holes are burned and probed by ramping the laser frequency using a double-passed AOM driven by a voltage-controlled oscillator. A fourth AOM is placed before the detector to block excitation pulses and selectively pass emitted echo signals or spectral hole scans. The maximum laser power at the sample is typically 100 mW for echo measurements and  $<$1 mW for hole burning. Optical transmission is detected using an amplified Si photodiode for hole burning or a photomultiplier for photon echoes. For two-pulse echo measurements, we employ pulse lengths of 200 ns for both pulses. See Ref. \cite{macfarlane1987a} for a further description of hole burning and photon echo measurement techniques.

\noindent
\textbf{\textit{Results--}} In the first experiment, we probe the entire absorption line of the $^3$H$_6$$\leftrightarrow$$^3$H$_4$ inhomogeneously broadened transition. Efficient multimode AFC memories require the optical transition to possess bandwidths that can accommodate interfacing with spectrally-multiplexed photon pairs. The time-bandwidth product $BT_{mem}$, where $B$ is the single-channel AFC bandwidth and $T_{mem}$ is the storage time, defines the maximum number of temporal modes accommodated by a single-channel AFC \cite{afzelius2009a}, while $\Gamma_{inh}/B$, where $\Gamma_{inh}$ is the inhomogeneous linewidth, approximately defines the maximum number of spectral modes accommodated \cite{sinclair2014a}. From spectral hole burning measurements and photon echo excitation measurements \cite{sun2012a} carried out across the inhomogeneous line, we measured a transition wavelength of 795.325 nm (vacuum), a full-width-half-maximum (FWHM) bandwidth of 56 GHz, and a peak absorption of 0.41 cm$^{-1}$. Additionally, we find uniform optical coherence properties over the inhomogeneous bandwidth, which allows its full use for spectral multiplexing. The 56 GHz bandwidth is larger than that observed in most high-quality RE-doped crystals \cite{sun2005a, macfarlane1987a,thiel2011a}. When employing a spectrally multimode AFC structure with single-channel width and a channel spacing of 100 MHz, Tm:YGG offers simultaneous storage of up to 280 spectral modes (we discuss later how to further increase this number). Furthermore, consolidating a Tm:YGG crystal into a monolithic impedance-matched cavity \cite{sabooni2013a,afzelius2010b,moiseev2010a} allows, in principle, increasing the memory efficiency arbitrarily close to unity.

\begin{figure}
\begin{center}
\includegraphics[width=\columnwidth]{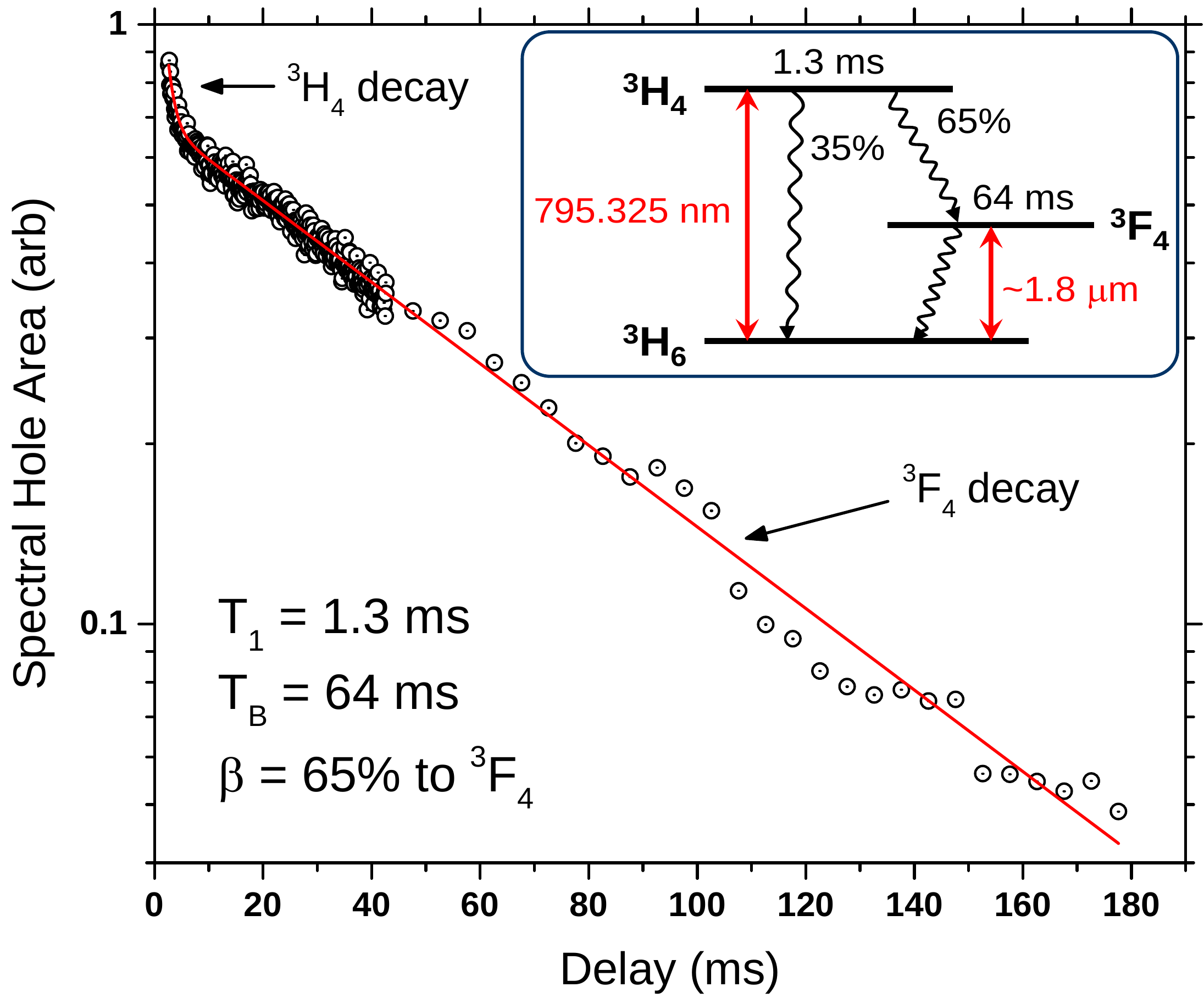}
\caption{\textbf{Transient spectral hole decay}. Hole burning measurements at 10 K reveal a $^3$H$_4$ lifetime of 1.3 ms and an exceptionally long $^3$F$_4$ lifetime of 64 ms. A relatively large probability $\beta$=0.65 of decay to the $^3$F$_4$ bottleneck level is measured. \textbf{Inset:} Simplified energy level diagram showing measured excited-state lifetimes, branching ratio, and transition wavelengths.
} 
 \label{fig:transdecay}
\end{center}
\end{figure}

In the second experiment, we perform time-resolved transient hole burning measurements to determine the $^3$H$_4$ excited-state lifetime and the $^3$F$_4$ metastable `bottleneck' lifetime. Provided an AFC can be tailored by optical excitation and subsequent relaxation of ions into a shelving level with lifetime $T_s$, the ratio $T_s / T_1$, where $T_1$ is the $^3$H$_4$ lifetime, must be much greater than unity to achieve high-fidelity storage \cite{saglamyurek2011a,lauritzen2010a}. An example spectral hole decay curve is plotted in Fig. \ref{fig:transdecay}, where we observe a double-exponential decay due to relaxation from the excited state and subsequent relaxation from the bottleneck level. This behavior is described by the relation
\begin{equation}
\Delta \alpha(t)= \Delta \alpha(0) [e^{-t/T_1}+\tfrac{\beta T_B}{2(T_B - T_1)} (e^{-t/T_B}-e^{-t/T_1})] .
\label{holedecay}
\end{equation}
The time-dependent hole depth $\Delta \alpha(t)$ decreases from its initial value of $\Delta \alpha(0)$ due to relaxation from the $^3$H$_4$ state with lifetime $T_1$, where a fraction $(1-\beta)$ of the population relaxes directly to the $^3$H$_6$ ground state and the other $\beta$ fraction of the population relaxes to the bottleneck level, which then decays with a lifetime $T_B$. The fit of the data using Eq. \ref{holedecay} is shown by the solid line in Fig. \ref{fig:transdecay}, indicating values of $T_1 = 1.3$ ms, $\beta = 65\%$, and $T_B = 64$ ms. The lifetime of the $^3$F$_4$ level is surprisingly long, and is one of the longest observed in solid state materials, particularly for RE ions at a crystallographic site without inversion symmetry. These results suggest a unique opportunity for AFC implementation where the high branching ratio and exceptionally long $^3$F$_4$ lifetime leads to the possibility of using an excited electronic level as a population reservoir for spectral tailoring (i.e. $T_B \equiv T_s$).

\begin{figure}
\begin{center}
\includegraphics[width=\columnwidth]{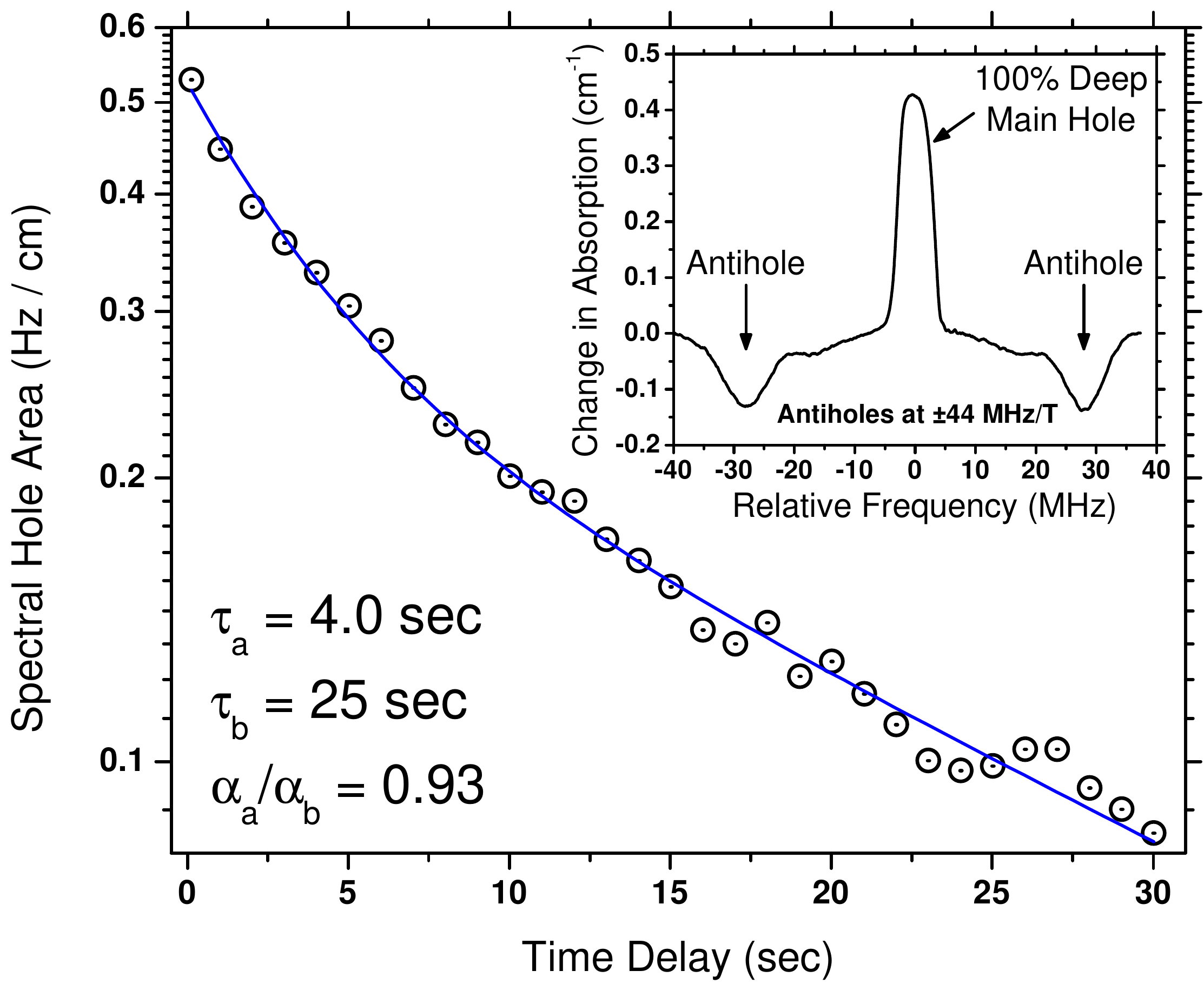}
\caption{\textbf{Persistent spectral hole decay measured at 2.0 K for a field of 272 G.} A double exponential decay is observed at low fields, possibly from two inequivalent subgroups of Tm$^{3+}$ ions. \textbf{Inset:} Measured persistent hole spectrum with an applied field of 503 G. A single pair of antiholes appear with a splitting of $\pm$44 MHz/T.
} 
 \label{fig:persistdecaylowB}
\end{center}
\end{figure}

In a third experiment, we examine the structure of persistent spectral holes burned with an applied magnetic field to investigate the $^{169}$Tm$^{3+}$ hyperfine states of the lowest energy $^3$H$_6$ and  $^3$H$_4$ levels \cite{macfarlane1987a}. Persistent spectral hole burning allows the maximum potential storage bandwidth of a single-channel AFC to be determined and also identifies any impact of atomic sub-structure on AFC quality (which, in turn, affects the storage efficiency) \cite{tittel2010a,afzelius2010a,saglamyurek2011a}. For $D_2$ point symmetry of the Tm$^{3+}$ site, the principle axes of the nuclear hyperfine interaction are oriented along the $<$001$>$, $<$110$>$, and $<$1-10$>$ sets of directions in the crystal lattice, resulting in six subgroups of magnetically inequivalent Tm$^{3+}$ ions in the lattice that have different local orientations. By specific choice of magnetic field and optical polarization directions, different subgroups of ions can be selectively probed and some subgroups can be made equivalent, significantly simplifying the material properties. In particular, when both the magnetic field and polarization are parallel to a $<$111$>$ direction, the optical field only interacts with half the Tm$^{3+}$ ions, all of which have the same hyperfine structure and optical Rabi frequency \cite{sun2000a}. For equivalent Tm$^{3+}$ ions in the lattice, each hole created in the spectrum can also result in one additional pair of sideholes and up to three pairs of antiholes due to optical transitions and population redistribution among the $^{169}$Tm$^{3+}$ nuclear hyperfine states. The depth of the sideholes and antiholes depends on the measurement timescale and the relative transition probabilities between hyperfine states of the $^3$H$_6$ and $^3$H$_4$ levels. However, if strong selection rules prevent optical transitions involving a change in the $^{169}$Tm nuclear spin state, the spectral hole structure is simplified so that only a single pair of antiholes appear, maximizing the bandwidth available for a single AFC channel prepared by optically pumping ions into the hyperfine states. A sample hole spectrum for a magnetic field of 0.64 T is shown in the inset of Fig. \ref{fig:persistdecaylowB}. The spectrum displays only a single pair of antiholes with a shift of $\pm$44 MHz/T relative to the main hole --  we observe no sideholes in either the persistent or transient hole burning measurements. We note that the relatively broad, laser-limited spectral hole width may obscure sideholes with small frequency shifts. Further measurements with a narrow-bandwidth laser or larger magnetic fields to probe any structure hidden within the main hole are required. However, we do not observe hyperfine modulation \cite{macfarlane1987a} in any of our photon echo decay measurements, further indicating a low probability for transitions involving a $^{169}$Tm nuclear spin flip.

We also measure lifetimes of the $^{169}$Tm nuclear spin states by observing the time-resolved persistent spectral hole area, as shown in Fig. \ref{fig:persistdecaylowB}, for a temperature of 2.0 K and a field of 272 G. While we would expect a single decay rate for these conditions, the observed decay is described by a double-exponential:
\begin{equation}
\Delta \alpha (t) = \Delta \alpha_a e^{-\frac{t}{\tau_a}}+\Delta \alpha_b e^{-\frac{t}{\tau_b}},
\label{persistdecay}
\end{equation}
where $\tau_a$ and $\tau_b$ are two different hole lifetimes and $\Delta \alpha_a$ and $\Delta \alpha_b$ are the relative amplitudes of the two population components present in the hole decay.  A fit using Eq. \ref{persistdecay} is shown by the solid line, giving lifetimes of 4.0 seconds and 25 seconds, respectively. The two components of the decay have nearly equal amplitudes with a ratio of 0.93, suggesting the presence of two subgroups of Tm$^{3+}$ in the lattice with identical optical properties but different nuclear spin relaxation times. Further studies are required to explore the nature of these two components and determine if they arise from perturbations of the local $D_2$ site symmetry, for example. At a much higher magnetic field of 0.64 T, the hole decays are well described by a single exponential. Measurements over timescales of 20 minutes yield an extrapolated lifetime of 6 hours. These lifetimes, in conjunction with the simple antihole structure, offer the possibility of tailoring AFCs with bandwidth $B = 100$ MHz under practical magnetic field strengths of a few Tesla \cite{tittel2010a,afzelius2010a,saglamyurek2011a}. Although the hyperfine splitting is an order of magnitude smaller than that in Tm$^{3+}$:LiNbO$_3$ \cite{sun2012a,sinclair2010a,thiel2010a,thiel2012a}, the AFC bandwidth in Tm:YGG with more than 1 T applied is similar to that offered in other materials with long coherence lifetimes such as Eu$^{3+}$:Y$_2$SiO$_5$ \cite{lauritzen2012a}.

\begin{figure}
\begin{center}
\includegraphics[width=\columnwidth]{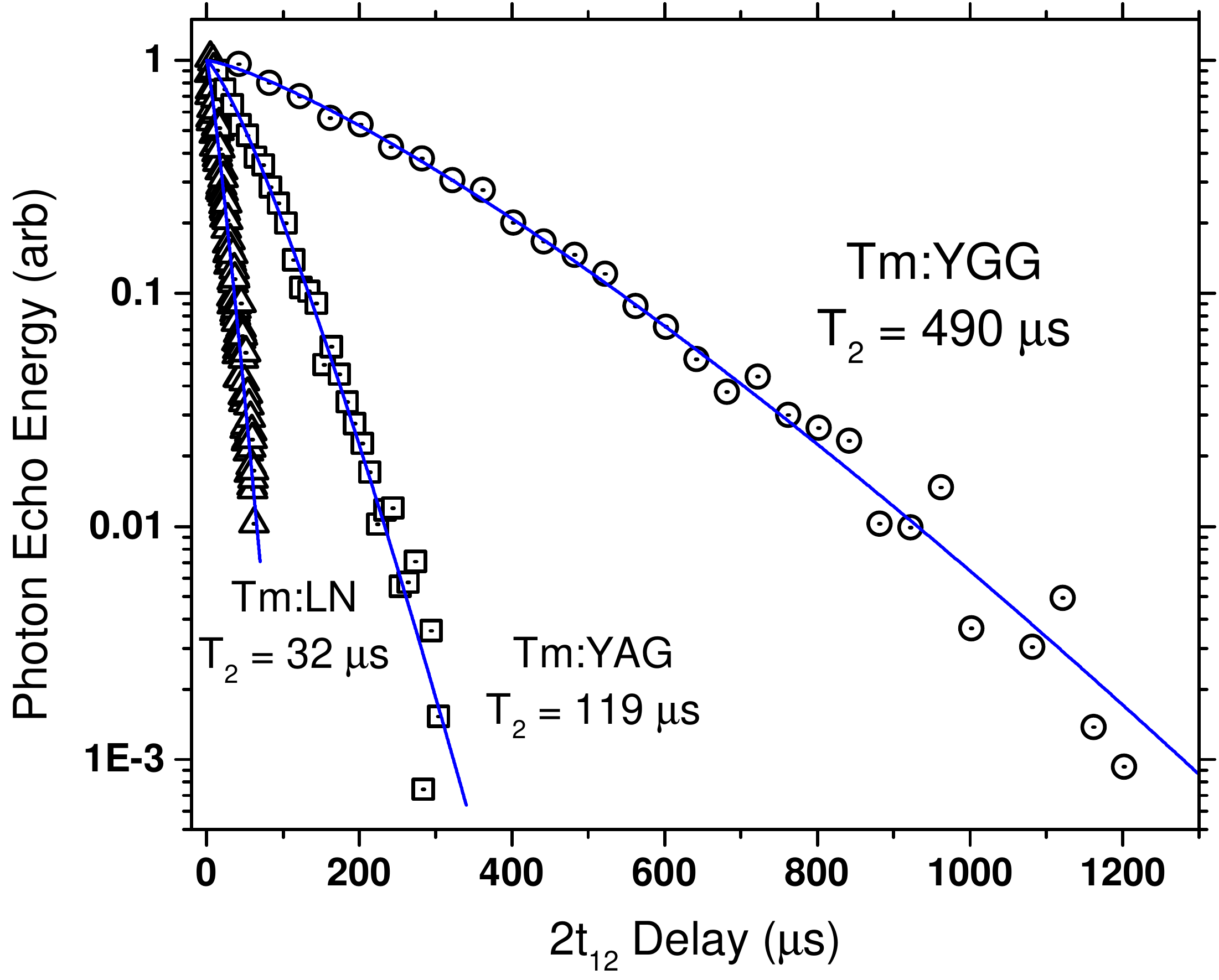}
\caption{\textbf{Two-pulse photon echo decays.} Measurements at 1.2 K and 454 G yield a coherence lifetime of 490 $\mu$s for 1\%Tm:YGG (circles). For comparison, echo decays are shown for Tm:YAG (squares) \cite{sun2005a} and Tm:LN (triangles) \cite{thiel2010a} at similar temperatures and fields.
} 
\label{fig:2pe}
\end{center}
\end{figure}
Finally, using two-pulse photon echo decay measurements, we probe the coherence lifetime, $T_2$, of the optical transition \cite{macfarlane1987a}, which bounds the maximum storage time for high-efficiency recall \cite{afzelius2009a,tittel2010a}. For this measurement, the laser is attenuated to minimize excitation-induced broadening caused by instantaneous spectral diffusion (ISD) \cite{thiel2014a}, the temperature is reduced to 1.2 K, and the magnetic field strength is increased to 454 G to inhibit spectral diffusion from nuclear spin flips or paramagnetic impurities. The photon echo decay under these conditions is plotted in Fig. \ref{fig:2pe}. To fit the non-exponential decay, we use the Mims expression 
\begin{equation}
I(t_{12}) = I_0 e^{-2(\tfrac{2 t_{12}}{T_2} )^x},
\label{mims}
\end{equation}
where $t_{12}$ is the delay between the two excitation pulses, $I_0$ is the initial echo intensity at $t_{12}=0$, $T_2$ is the time at which the echo intensity decays to $I_0 / e^2$, and $x$ is an empirical parameter that describes the decay shape \cite{mims1968}. We find a coherence lifetime $T_2$  of 490 $\mu$s -- the longest ever observed for optical transitions of Tm$^{3+}$ ions in a solid \cite{sun2005a,thiel2011a}. For comparison, the longest reported echo decays for the widely used materials 0.1\% Tm$^{3+}$:Y$_3$Al$_5$O$_{12}$ (Tm:YAG) \cite{sun2005a} and 0.1\%Tm$^{3+}$:LiNbO$_3$ (Tm:LN) \cite{thiel2010a} at similar temperatures ($< 2$ K) and applied field strengths ($>$ 200 G) are also shown in Fig. \ref{fig:2pe}, illustrating the significantly longer coherence lifetime provided by Tm:YGG. Note as well Pr$^{3+}$:Y$_2$SiO$_5$ and Eu$^{3+}$:Y$_2$SiO$_5$, other rare-earth-ion doped crystals that are currently being investigated for quantum state storage \cite{rielander2014a,jobez2014a,lauritzen2012a}, which offer comparably long optical coherence lifetimes up to 374 $\mu$s \cite{equall1995a} and 2.6 ms \cite{equall1994a}, respectively. The decay in Tm:YGG exhibits a weakly non-exponential shape with $x=1.35$, indicating that the coherence lifetime is limited by spectral diffusion likely resulting from perturbations due to interactions with fluctuating $^{69}$Ga and $^{71}$Ga nuclear spins in the host lattice. Since spectral diffusion magnitude and dynamics depend on temperature, magnetic field strength and orientation, and perhaps even crystal quality, we expect that optimization of the material and experimental conditions could further increase the coherence lifetime. The measured coherence lifetime of 490 $\mu$s establishes the maximum storage time with 90\% recall efficiency to be $T_{mem}=$ $274$ $\mu$s assuming rectangular-shaped AFC teeth; in the multiplexed quantum repeater architecture \cite{sinclair2014a}, this yields an elementary fiber-link length of $l_{link}=$ 55 km -- an encouraging result \cite{afzelius2010b,moiseev2010a,bonarota2010a}. However, additional studies of spectral diffusion over millisecond timescales are needed to determine the time-dependent decrease in coherence lifetime and the resulting AFC storage times that may be achieved in practice.

\noindent
\textbf{\textit{Conclusion--}} Our results suggest that Tm:YGG is the first, or, at least, the first Tm$^{3+}$-doped, material that satisfies the stringent combination of requirements for AFC-based broadband quantum memories needed for the quantum repeater proposed by Sinclair et al. \cite{sinclair2014a}. Among its many desirable properties, the convenient 795 nm wavelength and possibility for using the $^3$F$_4$ bottleneck level for population storage distinguishes Tm:YGG from other promising materials such as Pr$^{3+}$:Y$_2$SiO$_5$ or Eu$^{3+}$:Y$_2$SiO$_5$. Furthermore, Tm:YGG features less spectral diffusion and ISD than other Tm$^{3+}$ materials such as Tm:YAG \cite{thiel2014a}. Yet, more detailed measurements are required to determine the material's full potential, i.e. if the properties can be further improved by using different magnetic field orientations or optimizing the material composition. For example, we expect that the inhomogeneous broadening may be readily increased to hundreds of Gigahertz by introducing static crystal strain by doping selected impurities -- an approach that has been successfully demonstrated for the related material Tm:YAG \cite{sun2005a} --  increasing the number of 100 MHz-wide spectral channels to well above 1000. However, detailed studies of additional factors, such as time-dependent increases in the homogeneous linewidth due to spectral diffusion, must be performed to determine the material capabilities for specific quantum memory implementations. In parallel, optical pumping strategies must be optimized, e.g. by taking advantage of stimulated emission or spin-mixing to reduce the impact of population trapped in the long-lived $^3$H$_4$ state \cite{lauritzen2008a}. Furthermore, technical developments such as laser stabilization and embedding of Tm:YGG in impedance-matched cavities must occur to create optical quantum memories that enable long-distance quantum communications via quantum repeaters.

\noindent
\textbf{\textit{Acknowledgements--}} The authors acknowledge support from Alberta Innovates Technology Futures (AITF), the National Engineering and Research Council of Canada (NSERC), the Defense Advanced Research Projects Agency (DARPA) Quiness program (contract no. W31P4Q-13-l-0004), and the National Science Foundation (NSF) award nos. PHY-1212462 and PHY-1415628. Any opinions, findings and conclusions or recommendations expressed in this material are those of the authors and do not necessarily reflect the views of DARPA or NSF. W.T. is a senior fellow of the Canadian Institute for Advanced Research (CIFAR).


\begin{thebibliography}{1} 

\bibitem{sangouard2011a}
	N. Sangouard, C. Simon, H. de Riedmatten, and N. Gisin, Rev. Mod. Phys. \textbf{83}, 33 (2011).

\bibitem{lvovsky2009a}
	A. I. Lvovsky, B. C. Sanders, and W. Tittel, Nat. Photon. \textbf{3}, 706 (2009).

\bibitem{bussieres2013a}
	F. Bussi\`eres, N. Sangouard, M. Afzelius, H. de Riedmatten, C. Simon, and W. Tittel, J. of Mod. Opt. \textbf{60}, 1519 (2013).

\bibitem{afzelius2009a}
	M. Afzelius, C. Simon, H. de Riedmatten, and N. Gisin, Phys. Rev. A \textbf{79}, 052329 (2009).

\bibitem{sinclair2014a}
	N. Sinclair, E. Saglamyurek, H. Mallahzadeh, J. A. Slater, M. George, R. Ricken, M. P. Hedges, D. Oblak, C. Simon, W. Sohler, and Wolfgang Tittel, Phys Rev. Lett. \textbf{113}, 053603 (2014).

\bibitem{tittel2010a}
	W. Tittel, M. Afzelius, T. Chaneli\`{e}re, R. L. Cone, S. Kr\"{o}ll, S. A. Moiseev, and M. Sellars, Laser \& Photonics Reviews \textbf{4(2)}, 244 (2010).

\bibitem{sabooni2013a}
	M. Sabooni, Q. Li, S. Kr\"oll, and L. Rippe, Phys. Rev. Lett. \textbf{110}, 133604 (2013).

\bibitem{sinclair2010a}
	N. Sinclair, E. Saglamyurek, M. George, R. Ricken, C. La Mela, W. Sohler, and W. Tittel, J. Lumin. \textbf{130}, 1586 (2010).

\bibitem{riedmatten2008a}
	H. de Riedmatten, M. Afzelius, M. Staudt, C. Simon, and N. Gisin, Nature \textbf{456}, 773 (2008).

\bibitem{afzelius2010a}
	M. Afzelius, I. Usmani, A. Amari, B. Lauritzen, A. Walther, C. Simon, N. Sangouard, J. Min\'a\ifmmode~\check{r}\else \v{r}\fi{}, H. de~Riedmatten, N. Gisin, and S. Kr\"{o}ll, Phys. Rev. Lett., \textbf{104} 040503 (2010).

\bibitem{hedges2010a}
	M. P. Hedges, J.J. Longdell, Y. Li, and M. J. Sellars, Nature \textbf{465}, 1052 (2010).

\bibitem{clausen2011a}
	C. Clausen, I. Usmani, F. Bussi\`eres, N. Sangouard, M. Afzelius, H. de Riedmatten, and N. Gisin, Nature \textbf{469}, 508 (2011).

\bibitem{saglamyurek2011a}
	E. Saglamyurek, N. Sinclair, J. Jin, J. A. Slater, D. Oblak, F. Bussi\`eres, M. George, R. Ricken, W. Sohler, and W. Tittel, Nature \textbf{469}, 512 (2011).

\bibitem{jin2013a}
	J. Jin, J. A. Slater, E. Saglamyurek, N. Sinclair, M. George, R. Ricken, D. Oblak, W. Sohler, and W. Tittel, Nat. Comm. \textbf{4}, 2386 (2013). 

\bibitem{rielander2014a} 
	D. Riel\"{a}nder, K. Kutluer, P. M. Ledingham, M. G\"{u}ndo\v{g}an, J. Fekete, M. Mazzera, and H. de Riedmatten, Phys. Rev. Lett. \textbf{112}, 040504 (2014).

\bibitem{jobez2014a}
	P. Jobez, I. Usmani, N. Timoney, C. Laplane, N. Gisin, M. Afzelius, \textit{arXiv:1404.3489}.

\bibitem{sun2005a} 
	Y. Sun, \textit{Spectroscopic Properties of Rare Earths in Optical Materials}, Springer Ser. in Mater. Sci., Liu, G.K. and Jacquier, B., Eds., Berlin, Heidelberg, New York: Springer, 2005, Ch. 7.

\bibitem{thiel2011a}
	C. W. Thiel, T. B\"{o}ttger, and R. L. Cone, J. Lumin. \textbf{131}, 353 (2011).

\bibitem{macfarlane1987a}
	R. M. Macfarlane and R. M. Shelby, in \textit{Spectroscopy of Solids Containing Rare Earth Ions}, edited by A. A. Kaplianskii and R. M. Macfarlane (North Holland, Amsterdam, 1987), Chap. 3.

\bibitem{afzelius2010b}
	M. Afzelius and C. Simon, Phys. Rev. A \textbf{82}, 022310 (2010).

\bibitem{moiseev2010a}
	S. A. Moiseev, S. N. Andrianov, and F. F. Gubaidullin, Phys. Rev. A \textbf{82}, 022311 (2010).

\bibitem{sun2012a}
	Y. Sun, C. W. Thiel, and R. L. Cone, Phys. Rev. B \textbf{85}, 165106 (2012).

\bibitem{lauritzen2010a}
	B. Lauritzen, J. Min\'{a}\v{r}, H. de Riedmatten, M. Afzelius, N. Sangouard, C. Simon, and N. Gisin, Phys. Rev. Lett. \textbf{104}, 080502 (2010).

\bibitem{sun2000a}
	Y. Sun, G. M. Wang, R. L. Cone, R. W. Equall, and M. J. M. Leask, Phys. Rev. B \textbf{62}, 15443 (2000).

\bibitem{thiel2010a}
	C.W. Thiel, Y. Sun, T. B\"{o}ttger, W. R. Babbitt, and R. L. Cone, J. Lumin. \textbf{130}, 159 (2010).

\bibitem{thiel2012a}
	C. W. Thiel, Y. Sun, R. M. Macfarlane, T. B\"{o}ttger, and R. L. Cone, J. Phys. B: At. Mol. Opt. Phys \textbf{45}, 124013 (2012).

\bibitem{lauritzen2012a}
	B. Lauritzen, N. Timoney, N. Gisin, M. Afzelius, H. de Riedmatten, Y. Sun, R. M. Macfarlane, and R. L. Cone, Phys. Rev. B \textbf{85}, 115111 (2012).

\bibitem{thiel2014a}
	C. W. Thiel, R. M. Macfarlane, Y. Sun, T. B\"ottger, N. Sinclair, W. Tittel and R. L. Cone, Laser Phys. \textbf{24}, 106002 (2014).

\bibitem{mims1968}
	W. B. Mims, Phys. Rev. \textbf{168}, 370 (1968).

\bibitem{equall1995a}
	R. W. Equall, R. L. Cone, and R. M. Macfarlane, Phys. Rev. B \textbf{52}, 3963 (1995).

\bibitem{equall1994a}
	R. W. Equall, Y. Sun, R. L. Cone, and R. M. Macfarlane, Phys. Rev. Lett. \textbf{72}, 2179 (1994).

\bibitem{bonarota2010a}
	M. Bonarota, J. Ruggiero, J. -L. Le Gou\"{e}t, and  T. Chaneli\`{e}re, Phys. Rev. A \textbf{81}, 033803 (2010).

\bibitem{lauritzen2008a}
	B. Lauritzen, S. R. Hastings-Simon, H. de Riedmatten, M. Afzelius, and N. Gisin, Phys. Rev. A \textbf{78}, 043402 (2008).

\end{thebibliography}
\end{document}